\documentclass[twocolumn,showpacs,prl]{revtex4}
\usepackage{graphics,german,amsmath,amssymb}

\usepackage{graphicx}
\usepackage{color}
\usepackage{bm}

\newcommand{\mvec}[1]{{\bf #1}}
\newcommand{\mmatrix}[1]{\underline{\underline{#1}}}

\definecolor{olive}{rgb}{0.0,0.5,0.0}
\definecolor{darkred}{rgb}{0.5,0.0,0.0}

\begin{document}

\title{High field level crossing studies  on spin dimers in the low dimensional quantum spin system  Na$_2$T$_2$(C$_2$O$_4$)$_3$(H$_2$O)$_2$ with T=Ni,Co,Fe,Mn}
\author{C. Mennerich$^{1}$}
\author{H.-H. Klauss$^{1}$}
\author{A.U.B. Wolter$^{1}$}
\author{S. S\"ullow$^{1}$}
\author{F.J.~Litterst$^{1}$}
\author{C. Golze$^{1,2}$}
\author{R. Klingeler$^{2}$}
\author{V. Kataev$^{2}$}
\author{B. B"uchner$^{2}$} 
\author{M. Goiran$^{4}$}
\author{H. Rakoto$^{4}$} 
\author{J.-M. Broto$^{4}$}
\author{O. Kataeva$^{5}$}
\author{D.J. Price$^{6}$}

\affiliation{$^{1}$Institut f\"ur Physik der Kondensierten Materie, TU Braunschweig, Mendelssohnstr.3, D-38106 Braunschweig, Germany}
\affiliation{$^{2}$Leibniz-Institute for Solid State and Materials Research IFW Dresden, P.O. Box 270116, D-01171 Dresden, Germany}
\affiliation{$^{4}$ Laboratoire National des Champs Magn\'{e}tiques Puls\'{e}s, 
31432 Toulouse Cedex 04, France}
\affiliation{$^{5}$ Arbuzov Institute of Organic and Physical Chemistry, 
RAS, 420088 Kazan, Russia}
\affiliation{$^{6}$ WestCHEM,Department of Chemistry, University of Glasgow, Glasgow, G12 8QQ, UK}
\date{\today }

\begin{abstract}

In this paper we demonstrate the application of high magnetic fields to study the magnetic properties of low dimensional spin systems. We present a case study on the series of 2-leg spin-ladder compounds Na$_2$T$_2$(C$_2$O$_4$)$_3$(H$_2$O)$_2$ with T~=~Ni, Co, Fe and Mn. In all compounds the transition metal is in the $T^{2+}$ high spin configuation. The localized spin varies from S=1 to 3/2, 2 and 5/2 within this series. The magnetic properties were examined experimentally by magnetic susceptibility, pulsed high field magnetization  and specific heat measurements. The data are analysed using a spin hamiltonian description. Although the transition metal ions form structurally a 2-leg ladder, an isolated dimer model consistently describes the observations very well. This behaviour can be understood in terms of the different coordination and superexchange angles of the oxalate ligands along the rungs and legs of the 2-leg spin ladder. All compounds exhibit magnetic field driven ground state changes which at very low temperatures lead to a multistep behaviour in the magnetization curves. In the Co and Fe compounds a strong axial anisotropy induced by the orbital magnetism leads to a nearly degenerate ground state and a strongly reduced critical field. We find a monotonous decrease of the intradimer magnetic exchange if the spin quantum number is increased.

\end{abstract}


\maketitle

\section{Introduction}

In recent years, the physical properties of low dimensional spin systems have attracted a lot of attention. For isotropic magnetic interactions described in the Heisenberg model and low spin quantum numbers the ground state properties are strongly influenced by quantum fluctuations and often pure quantum ground states are found in a macroscopic system \cite{Takahashi99,Kluemper00,RiceScience96,DagottoPhyRevB92,DagottoReview}. Therefore these systems are ideal model systems for quantum mechanics. Low dimensional spin systems can be realized by linking transition metal via via organic molecules. Using the rich variety of organic ligands on the transition metal complex the dimensionality and strength of the magnetic interaction can be controlled.

Within this Springer series of lecture notes in physics a recent volume is devoted to high magnetic field studies in physics \cite{highfieldbook}. In that book several contributions describe the physics of one dimensional magnets in high magnetic fields \cite{BroholmHFB,RiceHFB} and electron spin resonance (ESR) on molecular magnets \cite{GatteschiHFB}. 
In this article we demonstrate the application of high magnetic fields to study the magnetic properties of spin dimers and determine the parameters describing the system.  The magnetic properties are examined experimentally by magnetic susceptibility, pulsed high field magnetization, high field ESR  and specific heat measurements.

We present a case study on a series of structural 2-leg spin ladders with different transition metal ions, namely Na$_2$T$_2$(C$_2$O$_4$)$_3$(H$_2$O)$_2$. This series consists of four isostructural compounds with magnetic ions T=Ni(II), Co(II), Fe(II) and Mn(II) with spin 1, 3/2, 2 and 5/2 respectively. One aim of this work is to elucidate the effect of a gradual increase of the spin multiplicity towards classical magnetism. In particular, the characteristic signatures of a strong single ion anisotropy in high field magnetization measurements are shown. We also discuss the dependence of the magnetic superexchange interaction strength on the number of 3d electrons, the influence of orbital moments and the validity of a description in the spin Hamiltonian model using isotropic Heisenberg exchange. 

In the first and second part, we describe the chemical synthesis and crystal structure of the samples and give a short overview of the spin hamiltonian used in the analysis. In the following four parts, we discuss the different compounds starting with the Ni(II) and Mn(II) compounds which have an orbital singlet ground state, followed by the Co(II) and Fe(II) compounds with an orbitally degenerate ground state. Finally, we compare the results obtained on the four samples.

\section{Synthesis and crystal structure}

The compounds Na$_2$T$_2$(C$_2$O$_4$)$_3$(H$_2$O)$_2$ with T~=~Ni, Co, Fe and Mn (we abbreviate the chemical structure with STOX for the general structure and SNOX, SCOX, SIOX and SMOX for the individual compounds) are synthesized in a hydrothermal reaction from solutions containing a very high sodium halide concentration \cite{DJPDalton00,MennerichPhyRevB06}. They occur as green crystals of SNOX, purple crystals of SCOX, yellow crystals of SIOX and white crystals of SMOX. Phase homogeneity and purity of all compounds were established by a combination of optical microscopy, powder X-ray diffraction and elemental analysis. The crystal size of the different compounds is micro-crystalline for the samples SCOX, SIOX and SMOX. Only for the SNOX compound single crystals up to 2 mg were grown in lower yielding reactions by slowing down the cooling rate, and reducing the concentration of Ni$^{2+}$ and (C$_2$O$_4$)$^{2-}$.

Single crystal X-ray structure determination reveals STOX to crystallize in the monoclinic space group P2$_1$/c (\#14) with slightly different crystallographic axes lengths below 2\%. The key features of the structure are the following: The T(II) ion  experiences a pseudo-octahedral coordination environment. It is coordinated in a \emph{cis} geometry by two chelating and crystallographically 
independent oxalate dianions. The remaining coordination sites are filled by a monodentate oxalate oxygen atom and a water molecule. T-O bond lengths lie within the range 2.0 to 2.1 \AA. We note that opposite pairs of O have similar T-O lengths, and that the O$\cdots$O separations between opposite pairs, for which the bond vectors are nearly directed along the crystallographic axes, are in the range of 4 \AA~  with a slightly bigger separation along the a-axes (fig. 1). The overall structure can be regarded as an anionic [T$_2$(C$_2$O$_4$)$_3$]$_n^{2n-}$ network with a 1-D ladder like topology (fig. 1), where the metal ions form the vertices and the bridging oxalate ions form the \textit{bonds}. The two oxalate ions bridge 3d ions in quite different ways. One oxalate forms a symmetric bis-chelating bridging mode (with a crystallographic inversion located at the C-C centroid) and links pairs of nickel ions (d$_{T\cdots T(rung)}$ $\approx$ 5.3 \AA) forming the rungs of the ladder. The second oxalate is unsymmetrically coordinated, chelating to one T(II) ion and forming a monodentate coordination to a second T ion, with 1,3-\textit{syn anti} geometry through the bridging carboxylate, this mode forms the ladder legs (d$_{T\cdots T(leg)}$ $\approx$ 5.8 \AA). These structural linkages are likely to provide the only significant pathways for magnetic superexchange.

\begin{figure}[ht]
    \includegraphics[width=0.9\columnwidth]{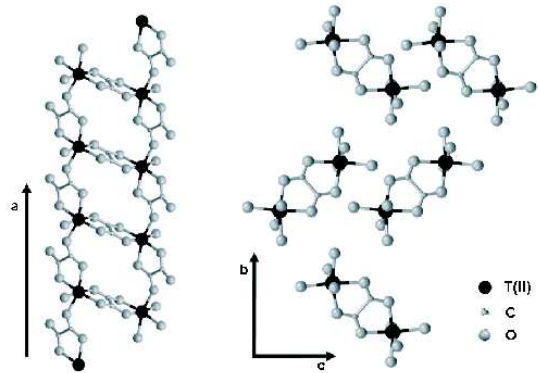}
    \caption{Crystal structure of  Na$_2$T$_2$(C$_2$O$_4$)$_3$(H$_2$O)$_2$. For clarity, Na and H$_2$ are not shown.}
    \label{Structure}
\end{figure}

\begin{figure}[ht]
    \includegraphics[width=0.9\columnwidth]{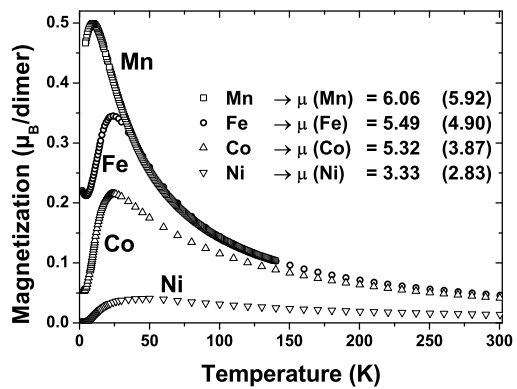}
    \caption{Temperature dependence of the magnetic susceptibility of powder samples of STOX . 
     The deduced high temperature magnetic moments (in units of $\mu_B$ in a Curie analysis) are shown in comparison with the spin only value calculated for the free ions (in brackets).}
		\label{mpowdervsT}
\end{figure}

The low dimensionality of the dominant magnetic exchange interaction in all compounds of this series is already evident from the magnetic susceptibility measured on powder samples between 2~K and 300~K (fig. \ref{mpowdervsT}). All compounds exhibit a Curie-Weiss behaviour at high temperatures,  a maximum between $\approx$ 10~K and 50~K and a strong decrease towards lower temperatures. This behaviour is typical for a system with a dominant antiferromagnetic interaction in less than three dimensions since no indications of static long range order are found down to 4~K. A Curie-Weiss analysis of the high temperature behaviour above 100~K yields effective magnetic moments close to the free ion high spin state for the Mn system only. While the data indicate moderate orbital contributions for the Ni and Fe systems, these contributions are strong for the Co system.

\section{Theoretical model}

To describe the magnetic properties we use the spin Hamiltonian approach. This approach neglects an explicit description of the orbital degrees of freedom  using pure spin coordinates. The orbital contributions treated as a perturbation lead to an anisotropic g-factor and an anisotropy of the spin orientation described by an anisotropy tensor $\mmatrix{D}$. The spin Hamiltonian approach is often valid for transition metal ions because their orbital moments are known to be quenched \cite{AB,Kahn,Gatteschi,Boca04}. It is usually working well for ions with orbitally non-degenerate ground states.

Taking into account the two different exchange pathway topologies, we assume a stronger magnetic interaction $J$ along the rungs than the magnetic interaction $K$ along the legs. This leads to an isolated dimer approximation where each dimer consists of two $S$=1, 3/2, 2 or 5/2 spins for Ni(II), Co(II), Fe(II) or Mn(II) respectively on the rungs of the ladder. Since all T ions on a ladder are crystallographically equivalent they share the same strength and orientation of the single ion anisotropy  $D$. Including the Zeeman energy in the external magnetic field, the spin Hamiltonian of the system in the dimer approximation is given by

\begin{equation}
    H  = J\, \mvec{S}_1 \mvec{S}_2
           + g \mu_B\,\mvec{B}(\mvec{S}_1 + \mvec{S}_2) 
           + \sum_{i=1,2}\, \mvec{S}_{i}\mmatrix{D}\mvec{S}_i. 
\end{equation}

Here, we assume that the magnetic field is applied along the $z$-axis, the main axis of the crystal field anisotropy tensor $\mmatrix{D}$. For pure axial anisotropy this tensor is  diagonal, defined as $D_{xx}$=~-~1/3~$D$,  $D_{yy}$=~-~1/3~$D$ and $D_{zz}$=~+~2/3~$D$. The effects of an in-plane anisotropy could be considered in the anisotropy tensor by the parameter $E$ with $E = \frac{1}{2} ( D'_{xx} - D'_{yy})  $. Usually $|D| \ge 3 E$ holds. Therefore in our study we neglect the effect of
a possible in-plane anisotropy. To calculate the energy levels for applied fields along different directions, we rotate the axis of the axial anisotropy tensor by an angle $\gamma$ about a perpendicular axis. Then the spin Hamiltonian can be expressed as

\begin{equation}
H  =   J\, \mvec{S}_1 \mvec{S}_2
           + g_\gamma \mu_B\,B(S_{1z}+ S_{2z}) 
           +\sum_{i=1,2}  \mvec{S}_i\mmatrix{U}_\gamma^T\mmatrix{D}\mmatrix{U}_\gamma\mvec{S}_i\label{Ham}
\end{equation}

with the rotation matrix $\mmatrix{U}_\gamma$ and an angle dependent g-factor g$_{\gamma}$.

The magnetization in units of $\mu_B$ per dimer can then be calculated by using the equation

\begin{eqnarray}
M_{dimer}&=&\frac{\partial F}{\partial B}=g\mu_B\frac{\frac{\partial Z}{\partial B}}{Z}\label{mag1}\\
&=&g\frac{\sum_{i}e^{E_i/T}\mvec{\Psi}_i(J,D,g,\gamma)\mmatrix{S_z}\mvec{\Psi}_i(J,D,g,\gamma)}{\sum_{i}e^{E_i/T}}\label{mag}
\end{eqnarray}

where $F=-k_BTlnZ$ is the free energy, Z is the partition function and $\mvec{\Psi}_i(J,d,g,\gamma)$ and $E_i$ are the eigenvectors and eigenvalues of equation (\ref{Ham}). The total magnetization $M_{tot}$ then consists of the main contribution (\ref{mag}) and comprises also a temperature independent term $M_0\,=\,M_{dia}\,+\,M_{vv}$ which includes a diamagnetic contribution $M_{dia}$ and a Van-Vleck paramagnetic susceptibility $M_{vv}$, as well as a Curie contribution $C/T$ with a Curie constant $C$ owing to paramagnetic impurities:

\begin{equation}
M_{tot} = M_{dimer} + C/T + M_0. \label{totalmag}
\end{equation}

To analyse the magnetization data, we developed a fit routine which numerically diagonalizes the Hamiltonian (\ref{Ham}), calculates the magnetization $M_{tot}$ using equation (\ref{totalmag}) and varies the parameters in the Hamiltonian to minimize the mean square deviation between the data and this model.

The isolated dimer approximation may be improved by introducing an effective exchange interaction $K$ along the legs treated in mean field approximation in the calculation of the magnetic susceptibility $\chi$:

\begin{equation}
\chi_{ladder} = \chi_{dimer}/(1+K\chi_{dimer})\label{mfSus}.\\
\end{equation}

Furthermore we calculate the magnetic specific heat by calculating the partition function Z for the Hamiltonian (\ref{Ham})

\begin{eqnarray}
c_{p,mag}  =  T \, \partial/\partial T \, (lnZ \,- \, 1/T \, (\partial(ln Z)/\partial\beta))=\\
=T \, \frac{\partial}{\partial T} \, (ln(\sum_{i}e^{E_i/T}) \,- \, \frac{1}{T} \, \frac{\partial(ln(\sum_{i}e^{E_i/T})}{\partial\beta})) \label{cp}
\end{eqnarray}

with $\beta~=~1/k_BT$, leading to an analytical function for a given angle $\gamma$ which can be used to analyse 
the experimental specific heat with a standard $\chi^2$ fit routine.

\section{$\mathrm{Na}_2\mathrm{Ni}_2$(C$_2$O$_4$)$_3$(H$_2$O)$_2$}

Antiferromagnetic S=1 chain systems are of particular interest in quantum magnetism since they show a non-magnetic singlet ground state with a spin excitation gap (Haldane chains) \cite{HaldanePhysRevLett83,EXPHaldanechains}. The physical properties of two coupled
S=1 chains have been studied theoretically  \cite{SenechalPhysRevB95,AllenPhysRevB00,TodoPhysRevB01, SatoPhysRevB05} but no experimental realization of a Haldane ladder system has been identified to date. Therefore we start our investigations with the Na$_2$Ni$_2$(C$_2$O$_4$)$_3$(H$_2$O)$_2$ (SNOX) compound. In this compound, the Ni(II) ions are in the $3d^8$ configuration with an orbitally non-degenerate ground state and a total spin of S~=~1 per ion.

\subsection{Magnetic Susceptibility}

We examined several single crystals in magnetic susceptibility measurements using a Quantum Design Magnetic Properties Measuring System (MPMS) in external fields of $B_{ex}$~=~2~T  and 5~T in the temperature range 2-300~K.  In these experiments the external field was oriented along the $a$-, $b$- and $c$-axes as well as along different intermediate angles with respect to the $a$-axis.

\begin{figure}[ht]
    \includegraphics[width=0.9\columnwidth]{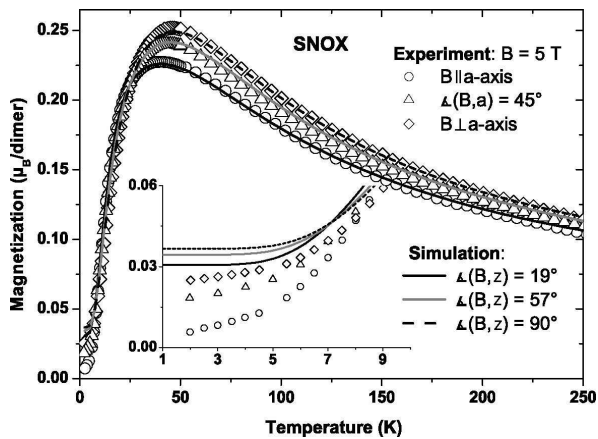}
    \caption{Temperature dependence of the magnetization of a SNOX single crystal at an external field of B$_{ex}$~=~5~T along the $a$-axis (circles) and at different angles with respect to the $a$-axis (triangles and squares). The solid line represents the fit as described in the text.}
		\label{mvsT}
\end{figure}

Fig. \ref{mvsT} shows the temperature dependence of the  magnetization along the $a$-axis and at different angles with respect to the $a$-axis in an external field of $B_{ex}$~=~5~T. Similar to the powder measurement presented in fig. \ref{mpowdervsT} and the single crystal measurements in  $B_{ex}$~=~2~T \cite{MennerichPhyRevB06} one can clearly see a pronounced maximum for all directions with a strong downturn below $\approx$ 50 K suggesting a non-magnetic spin-singlet ground state. This behavior in general is expected as well for an isotropic two-leg spin ladder as for the two limiting cases, an isolated $S$~=~1 Haldane chain or a system of antiferromagnetically coupled dimers \cite{HaldanePhysRevLett83,DagottoPhyRevB92,AllenPhysRevB00,SenechalPhysRevB95,TodoPhysRevB01,SatoPhysRevB05}.

The inset of fig. \ref{mvsT} shows the behavior at very low temperatures where the magnetization along a is smaller and approaches zero for T $\rightarrow$ 0 in contrast to directions transverse to a. Since the $a$-axis is close to the axial anisotropy axis as we will see below this indicates an easy axis of the Ni spin moments. The difference between experiment and theory at very low temperatures is due to an overestimated temperature independent Van Vleck contribution in the theoretical fit. 

To analyse the magnetization data, we performed a combined fit of $M_{tot}$ for $B_{ex}$~=~5~T along all measured directions using an isotropic magnetic exchange coupling constant $J$ and a fixed absolute value of the axial anisotropy of $|D|$~=~11.5~K. This anisotropy value was determined by the S=1 zero field splitting measured by ESR as described below. From these measurements
we obtain an intradimer coupling constant of $J$~=~44~K, an interdimer coupling constant of $K$~=~0~K and $g$-values of $g_a$~=~2.215, $g_{45}$~=~2.300~ and $g_{90}$~=~2.330~. The fitted angles to the anisotropy axis are $\gamma_a$~=~19$^\circ$, $\gamma_{45}$~=~57$^\circ$ and $\gamma_{90}$~=~90$^\circ$. In addition a temperature independent contribution $\chi_0\,=\,0.03\,\mu_B$/dimer has been used. The fit results are shown in fig. \ref{mvsT} as solid lines. They are in good agreement with the results obtained in \cite{MennerichPhyRevB06} for $B_{ex}$~=~2~T. 

\begin{figure}[ht]
    \includegraphics[width=0.9\columnwidth]{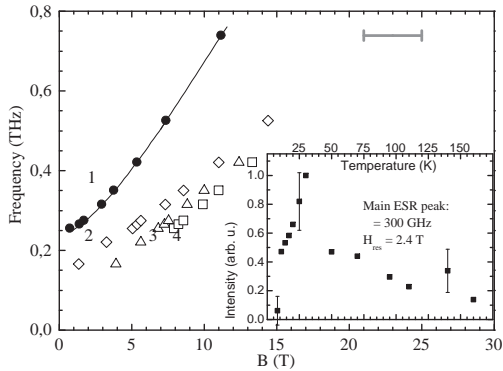}
    \caption{Main panel: Frequency $\nu$ vs. resonance field $B^{res}$ dependences (branches) of the strongest line 1 and weak satellites 2,3 and 4 comprising the
ESR spectrum of SNOX. Note the intercept of branch 1 with the frequency axis which yields the magnitude of the anisotropy parameter $D$. Solid line
is a theoretical fit according to Hamiltonian (2) with $|D| = 11.5$~K. Dashed lines are a linear approximation of weak satellite branches. Grey bar
indicates the spread of the overlapping satellite resonances in the high field regime. Inset: $T$-dependence of the strongest ESR line 1. Note its
similarity to the $T$-dependence of the static magnetization.}
		\label{ESR}
\end{figure}

To determine the absolute value of the axial anisotropy constant $D$ we have performed tunable high-field ESR measurements of SNOX at frequencies up
to $\nu\sim 1$~THz in magnetic fields up to $B\sim 40$~T. Details of experimental set-ups can be found in Ref.~\cite{Golze06}. A complex spectrum comprising a
main line and a number of weak satellites has been observed. The ESR-intensity shows a thermally activated behavior similar to the $T$-dependence of
the static magnetization (see fig.~\ref{ESR}, inset) thus ensuring that the ESR response is determined by the bulk Ni-spins. The frequency vs.
magnetic field dependence of the ESR modes is shown in the main panel of fig.~\ref{ESR}. In our analysis of the ESR data we focus the attention on
the strongest resonance line 1 which $\nu(B)$-dependence exhibits the intercept with the frequency axis at $\nu_0\,\approx\,239$\,GHz (details of the
ESR analysis can be found in Ref.~\cite{MennerichPhyRevB06}). This intercept implies a finite energy gap $\Delta\,=\,\nu_0h/k_B\,=\,11.5$\,K for this resonance excitation.
This zero field gap can be straightforwardly identified with the zero field splitting of the first excited $S = 1$ triplet state of the dimer and
gives directly the magnitude of the axial anisotropy parameter $\Delta = |D| = 11.5$~K.

The knowledge of $\gamma_a$, $\gamma_b$ and $\gamma_c$ \cite{MennerichPhyRevB06} allows a determination of the orientation of the
anisotropy axis in the crystal. It is oriented parallel to the connecting line of the opposite nearest neighbor oxygen ions along the $a$-axis, tilted by 18$^\circ$ with respect to the crystallographic $a$-axis mainly towards the b axis \cite{MennerichPhyRevB06}.
Note, that analyzing the data with a negative $D$~=-11.5~K leads to a similar good agreement between the model and the data. However, the positive $D$ value is strongly supported by the high field magnetization data described in the next section.

\begin{figure}
    \includegraphics[width=0.9\columnwidth,clip]{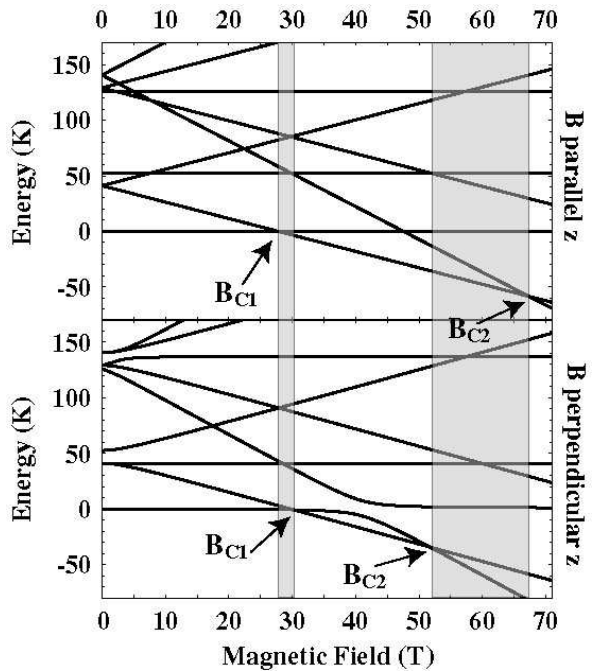}
    \caption{Relative energies of the spin states, calculated for the magnetic field applied parallel and perpendicular to the $z$~axis of the
      uniaxial anisotropy tensor: the 
      %Schwerpunkt des
      triplet $S\,=\,1$ and the quintet $S\,=\,2$ are well separated from the $S\,=\,0$ ground state by an activation energy of
      $|J|\,=\,44$\,K and $3\,|J|$, respectively. Both multiplets exhibit a zero field splitting due to crystal field anisotropy. The shaded
      areas mark the field ranges of ground state level crossings around $29$\, and $60$\, T in powder magnetization measurements (for details
      see text).}
\label{States}
\end{figure}

The relative energies of the spin states of SNOX calculated in the framework of the Hamiltonian (\ref{Ham}) with the parameters yielding the
best fit to the temperature dependent magnetization are plotted in fig.~\ref{States} for magnetic fields oriented parallel and perpendicular to the main axis of the anisotropy tensor. One can see that at zero field the spin singlet ground state $S\,=\,0$ is well separated from the $S\,=\,1$ triplet and the $S\,=\,2$ quintet state. The separation energy between singlet and triplet corresponds to the coupling constant $J$, whereas that between singlet and quintet corresponds to $3J$. The splitting of the excited $S\,=\,1$ and $S\,=\,2$ states in zero magnetic field is induced by the anisotropy $D$. Both triplet and quintet levels split with increasing magnetic field due to the Zeeman effect. The interplay  of the zero field splitting and the Zeeman splitting in the determination of the high field properties of SNOX will be discussed in detail in the next section.

\subsection{High field magnetization} \label{SNOXhighfield}

As can be seen from the Breit-Rabi diagram of the spin states of SNOX (fig.~\ref{States}) the splitting of the states in a magnetic field yields a
level crossing of the ground state with the lowest triplet state $\left|\,-\,\right>$ at a field $B_{C1}$ and a second level crossing of the
$\left|\,-\,\right>$ state with the $\left|\,-\,-\,\right>$ quintet state at field $B_{C2}$ with $B_{C1}\,<\,B_{C2}$. This leads to a step-like behavior in high field magnetization measurements. If the field is applied along the $z$~axis, the critical field $B_{C1}$ can be estimated using the equation $B_{C1}\,=\,(J-D/3)/(g\mu_B/k_B$), depending strongly on the sign of $D$. For a field perpendicular to the local anisotropy axis, a negative $D$ (leading to higher critical fields for $B\,\parallel\,z$) pushes the critical field to lower fields and vice versa for a positive $D$. Since in powder measurements the perpendicular situation dominates the spectrum of the spin states, the field dependent magnetization, among others, can determine the sign of the anisotropy $D$.

\begin{figure}[ht]
    \includegraphics[width=0.9\columnwidth]{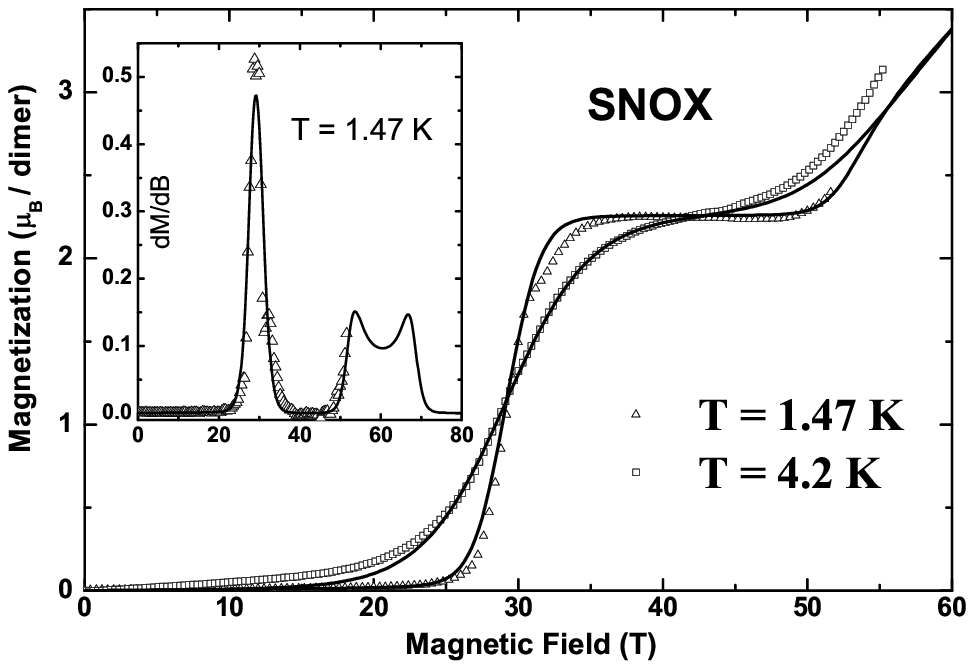}
    \includegraphics[width=0.9\columnwidth]{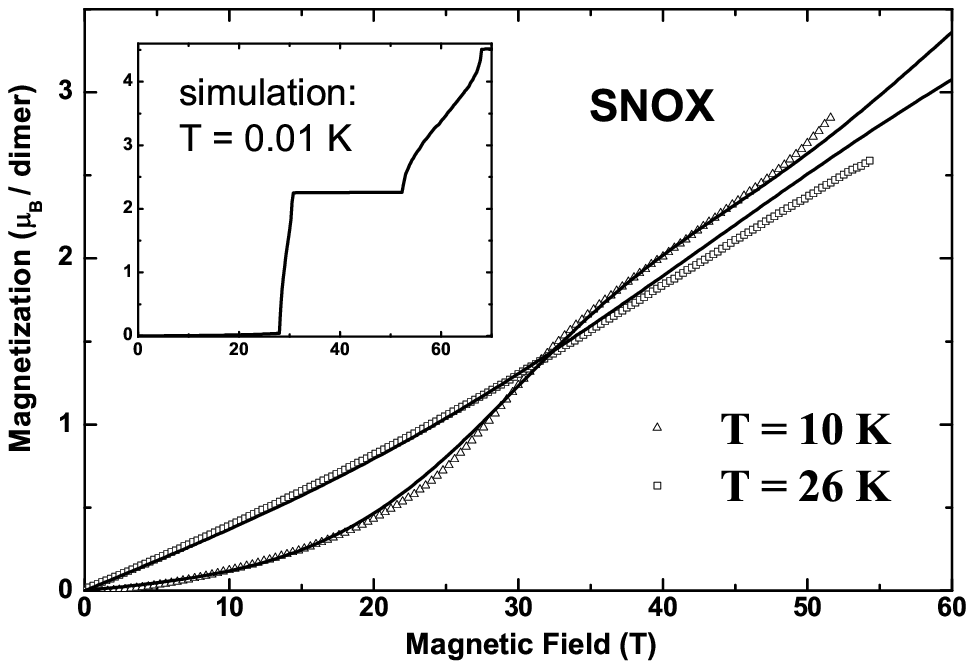}
    \caption{High field magnetization of a SNOX powder sample. Upper panel: Measurements at 1.47~K and 4.2~K, the inset shows the differential magnetization $\partial M / \partial B$ at 1.47~K. Lower panel: measurements at 10~K and 26~K, the inset shows the simulations for T~=~0.01K up to 80 T field strength. The solid line represents the simulation as described in the text.}
		\label{mvsB}
\end{figure}

The critical field strengths $B_{Ci}$ depend on the magnetic exchange $J$.  An exchange energy of $J$=44~K corresponds roughly to $B_{C1}$=30~T for $B_{C1}$. Therefore high magnetic fields are needed to observe the magnetization steps. We performed measurements at several temperatures (1.47~K, 4.2~K, 10~K and 26~K) on a powder sample with a mass of~23.2~mg in magnetic fields up to 55~Tesla. The results are shown in fig. \ref{mvsB}. In  the low temperature experiments the first magnetization step is clearly seen, whereas it smoothes at higher temperatures due to the thermal averaging process. The critical field $B_{C1}$ can be obtained from the derivation of the magnetization dM/dB
for $T$~=~1.47~K (see inset of the upper panel of fig. \ref{mvsB}), $T$~=~4.2~K and $T$~=~10~K. At all temperatures a sharp peak is observed at 29~Tesla. Additionally a second increase of the magnetization at much higher fields of about 50~Tesla can be anticipated. This step is clearly observed at $T$~=~1.47~K and $T$~=~4.2~K and strongly broadened in the 10~K measurement. At 26~K the magnetization steps are not observed and a linear field dependence of the magnetization is found.

The solid lines in fig.\ref{mvsB}  describe simulations using the dimer model with the calculated g values $g_\parallel$~=~2.201 (see above) and $g_\perp$~=~$g_{90}$. For the simulation of a powder measurement in this case a weighted average over a full set of angles between the anisotopy axis and the field direction from 0$^\circ$ to 90$^\circ$ in steps of 1$^\circ$ is performed with varying g from $g_\parallel$ to $g_\perp$. Since high magnetic fields lead to strong nonlinearities in
the magnetization curve this approach is more realistic than the two step average $M$~=~($M_\parallel$+2~$M_\perp$)/3.  
For a good description of the high field magnetization we need to use a coupling constant of $J$~=~44.5~K instead of $J$~=~44~K. A small deviation is found in the field range of 30~-~33~T in the $T$~=~1.47~K measurement, which may be caused by a little shift of the sample induced by the external field. 

The simulations for a negative $D$~=~-11.5~K did not describe the data reasonably well for any $J$. In this case the simulated magnetization step is clearly broader than the step of the experimental data set. The enhanced broadening of the magnetization step for $D$~=~-11.5~K can be explained by the resulting switching of the m~=~0 state with the $|m|$=1 states which leads to different level crossing field strengths $B_{C1}$.

From the differential magnetization $\partial M / \partial B$ we can determine the critical field $B_{C1}$ of the powder sample to $B_{C1}$~=~29~T. In our simulations, we get $B_{C1}$~=~27.8~T for $B$ parallel to the magnetic anisotropy axis  and $B_{C1}$~=~30.2~T for B perpendicular to the magnetic anisotropy axis, resulting also in $B_{C1}$~=~29~T in a powder average. Notice, that the powder averaging leads to a broadening of the magnetization steps. For the second step, we get $B_{C2}$~=~67.3~T and $B_{C2}$~=~51.1~T for B parallel and perpendicular, respectively.  This difference leads in a powder sample  at the lowest temperature of our experiment, $T$~=~1.47~K, to a continuous double sigmoid-like increase of the magnetization in the field range of 51~-~67 Tesla, of which only the first increase is observed in the measurement.

\subsection{Specific Heat}

\begin{figure}[ht]
\includegraphics[width=0.9\columnwidth]{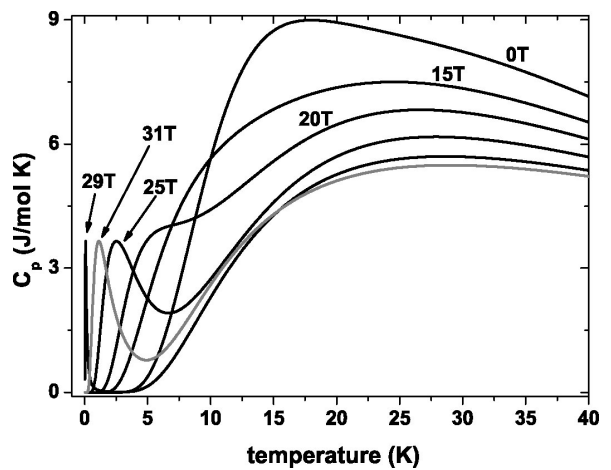}
\caption{Calculated temperature dependence of the magnetic specific heat of SNOX for high external fields.}
\label{cp-sim}
\end{figure}

As shown in fig. \ref{States} the energy gap~$\Delta$ between the singlet ground-state and the $\left|\,-\,\right>$ state of the magnetic S~=~1 triplet is reduced by an external field from $\Delta$$_{0T}$~$\approx$~41.1~K to zero for a field perpendicular to the anisotropy axis at a level crossing field of B$_{C1}$~$\approx$~30.2~T.  Then at higher field the gap increases with further increasing field until the $\left|\,--\,\right>$ state of the S~=~2 triplet anti-crosses the S~=~0 singlet at $\approx$~42~T . At higher fields the gap decreases again towards zero at the second ground state change at B$_{C2}$~$\approx$~51.3~T. This behaviour can be directly observed in the calculated magnetic specific heat according to equation (\ref{cp})(fig. \ref{cp-sim}). At low fields only one broad maximum is found. Increasing the field leads to a decrease of the gap. Approaching B$_{C1}$ a low temperature peak is separated from the main component. At the critical field this low temperature peak sharpens and shifts towards zero temperature whereas the maximum of the main component shifts to slightly higher temperatures. For fields above B$_{C1}$ the low temperature peak shifts to higher temperatures. 

\begin{figure}[ht]
\includegraphics[scale =0.9]{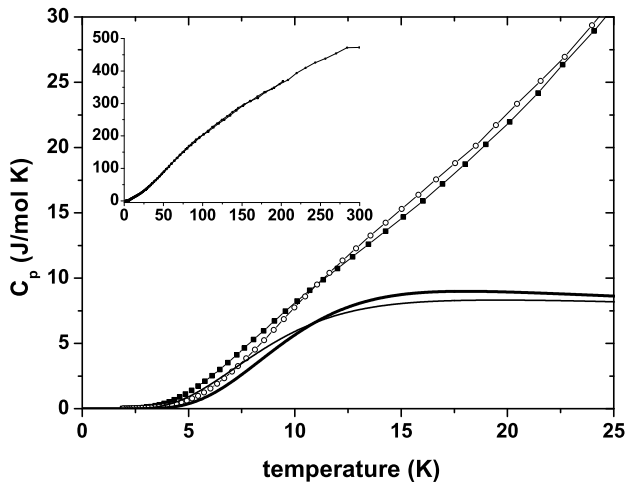}
\caption{Temperature dependence of the specific heat of a SNOX single crystal (2.04~mg) at zero field (open circles) and B$_{ex}$~=~9~T (filled squares) compared with the calculated magnetic specific heat  for zero field (thick line) and B$_{ex}$~=~9~T (thin line). The inset shows the measurements on an extended temperature range.}
\label{cp-meas}
\end{figure}

Measurements of the specific heat c$_p$(T) were performed on a single crystal of mass 2.04~mg using a Quantum Design PPMS in zero-field and an external field of B~=~9~T perpendicular to the a-axis in the temperature range 2~-~300~K (fig.\ref{cp-meas}). The measurements differ only in the low temperature regime where the magnetic specific heat is dominant \cite{MennerichJMMM06}.

\begin{figure}[ht]
\includegraphics[scale =0.9]{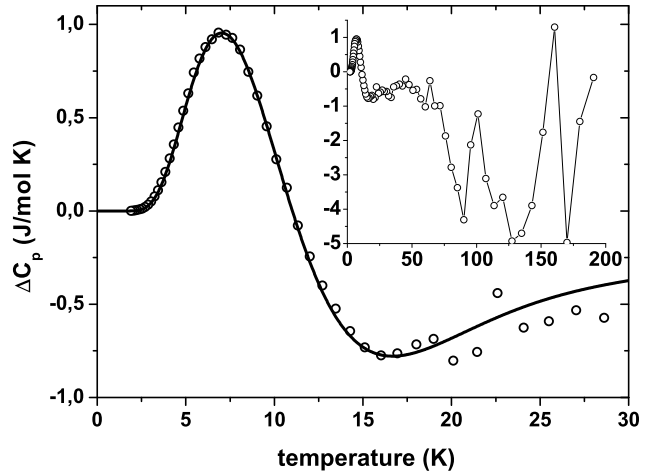}
\caption{Specific heat difference c$_p$~(9T)~-~c$_p$~(0T) of SNOX. The inset shows the difference in an extended temperature range.}
\label{cpdiff-fit}
\end{figure}

Due to the large coupling constant of the dimer, the magnetic contribution to c$_p$(T) is strongest in the temperature range around T~$\approx$~15~--~ 50~K. Since the Debye temperature T$_D$ of such metal-organic compounds is typically in the range of 100~-~200~K, the magnetic contribution to c$_p$ cannot be extracted with a standard T$^3$-fit for the lattice contribution which is possible only for T~$\ll$~T$_D$.  By analyzing the difference of both measurements c$_p$~(9T)~-~c$_p$~(0T) as shown in fig.\ref{cpdiff-fit} we subtract the phonon contribution. A fit using the isolated dimer model with variable parameters J and g in the temperature range T~$<$~19~K results in a very good agreement for a coupling constant J~=~43.7(5)~K and a g-value of 2.41(5) \cite{MennerichJMMM06}. 

In conclusion, the specific heat measurements on SNOX are consistent with the susceptibility and high field magnetization measurements. All experiments show that the isolated dimer model in the spin hamiltonian approach provides a very good description of the system. This can be understood considering the different coordination and superexchange angles of the exchange mediating oxalate molecule along the rungs and legs of the ladder. A detailed discussion is presented below. The derived parameter values for $D$, $J$ and the $g$-factor are typical for Ni(II) in a pseudo-octahedral environment and  oxalate-bridged dimers \cite{Ni-sia-1,Ni-sia-2,Ni-sia-3,Ni-sia-4,Vitoria03,escuerICA95,Ni-dim-1}.

\section{$\mathrm{Na}_2\mathrm{Mn}_2$(C$_2$O$_4$)$_3$(H$_2$O)$_2$}
\subsection{Magnetic Susceptibility}

In Na$_2$Mn$_2$(C$_2$O$_4$)$_3$(H$_2$O)$_2$ the Mn(II) ($3d^5$) configuration results in an orbitally non-degenerate ground state  and a total spin of S~=~5/2 per ion. This leads to a very specific behaviour compared with the Ni(II), Co(II) and Fe(II) ions: Whereas for the ions with a more than half-filled 3d shell the orbital magnetic moments are quenched and have to be considered as a perturbation, the orbital angular momentum L for the $3d^5$ configuration is zero following Hund's rules. Therefore we can neglect the anisotropy term and the Hamiltonian simplifies to
\begin{equation}
    H  = J\, \mvec{S}_1 \mvec{S}_2
           + g \mu_B\,\mvec{B}(\mvec{S}_1 + \mvec{S}_2). \label{isotropHam}
\end{equation}

\begin{figure}[ht]
    \includegraphics[width=0.9\columnwidth]{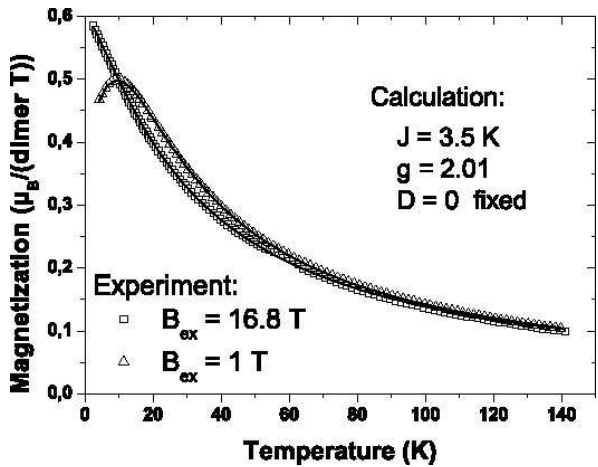}
    \caption{Temperature dependence of the magnetization of a SMOX powder sample at external fields of B$_{ex1}$~=~1~T (triangles) and B$_{ex1}$~=~16.8~T (squares). The 16.8~T measurement is divided by the field. The solid line represents the fit as described in the text.}
		\label{MnmvsT}
\end{figure}

We measured a powder sample of mass~24.8~mg in a home built vibrating sample magnetometer \cite{DissvonMarkus} in external fields of $B_{ex}$~=~1~T  and 16.8~T in the temperature range 2-300~K. The measurements are shown in fig.(\ref{MnmvsT}). For $B_{ex}$~=~1~T a clear maximum 
is observed at around 10~K  whereas for $B_{ex}$~=~16.8~T the magnetization increases monotonously with decreasing temperature indicating that at this field strength the antiferromagnetic intradimer interactions are suppressed by the strong Zeeman interaction of the individual spins in the external field.

To analyse the magnetization data, we performed a combined fit of $M_{tot}$ for $B_{ex}$~=~1~T and $B_{ex}$~=~16.8~T using an isotropic magnetic exchange coupling constant $J$ and an averaged g-factor. Since without anisotropy the magnetizations for applied external fields parallel and perpendicular to the $z$-axis are identical we performed an analysis without powder averaging.
We obtained an intradimer coupling constant of $J$~=~3.5~K and a $g$-value of $g$~=~2.01. The fit is done without a temperature independent contribution and without considering paramagnetic impurities. The results are shown in fig. \ref{MnmvsT} as solid lines. To prove the validity of the isotropic model (i.e. the $^6S_{5/2}$ ground state) we performed an analysis of the $B_{ex}$~=~1~T measurement including an anisotropy D and an anisotopic g-factor in a powder average. This approach did not improve the quality of the fit. The same holds for including an interdimer coupling K on the mean field level.

The different behaviour of the 1~T and the 16.8~T magnetization curves can be explained considering
the energy levels of the system. The Breit-Rabi diagram of SMOX shows 6 multiplets, one for each 
spin between S~=~0 and S~=~5. These multiplets are degenerate for B~=~0 reflecting the absent anisotropy D. In an external magnetic field the multiplets split due to the Zeeman interaction and five ground state level crossings are observed: for J~=~3.5~K and g~=~2.01 at approximately 2.6, 5.2, 7.8, 10.4 and 13~T. A non-magnetic ground state (i.e. $\partial E/\partial B~=~0$) is only found for fields below 2.6~T. At higher fields a magnetic ground state exists with a finite magnetization for $T\rightarrow 0$ which increases steplike at the level crossings. The different ground states can be observed in the low temperature behaviour of the temperature dependent magnetization curves. 
The temperature dependence of the magnetization for fields below 2.5~T  shows a strong downturn to zero reflecting the depopulation of the magnetic levels whereas for higher fields a finite value of 2, 4, 6, 8 and 10~$\mu_B$ per dimer is found for the different field ranges. For the highest field regime above 13~T, the magnetization increases monotonously with decreasing temperature to the maximum value of 10~$\mu_B$ per dimer.

\subsection{High field magnetization} \label{SMOXhighfield}

\begin{figure}[ht]
    \includegraphics[width=0.9\columnwidth]{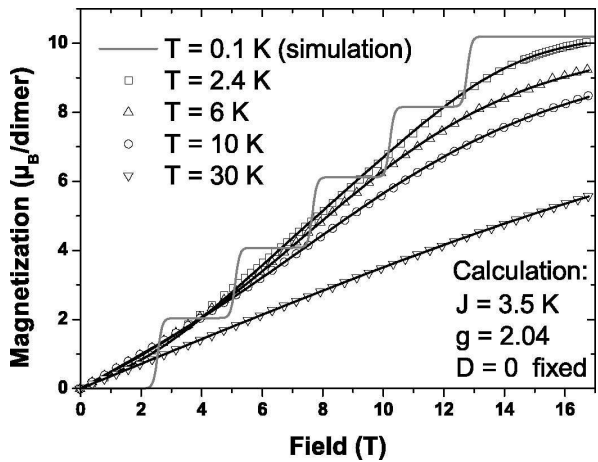}
    \caption{High field magnetization of a SMOX powder sample. The figure shows measurements at 2.4~K, 6~K, 10~K and 30~K. The black solid lines represents the fit as described in the text and the grey solid line is a calculated magnetization for T~=~0.1~K.}
		\label{MnmvsB}
\end{figure}

Field dependent magnetization measurements were also performed in the vibrating sample magnetometer on  a powder sample at temperatures 2.4, 6, 10 and 30~K in fields up to 16.8~T. The measurements are shown in fig.(\ref{MnmvsB}).

The analysis is done using the Hamiltonian (\ref{isotropHam}). The eigenvalues can be calculated directly using 
\begin{equation}
    E_n  = J/2~(S(S+1)-(S_1(S_1+1)-S_2(S_2+1))\\
    -g~mu_B~B~(S_{1z}+S_{2z}) \label{evisotrop}
\end{equation}
were the basis $\left|\,S~m_s\,\right>$ is given by total spin $S~=~S_1~+~S_2~=0...5$ with magnetic quantum numbers $m_s$ from -S to S leading to 36 eigenvalues.

This basis allows to calculate the exact eigenvalues with parameters $J$, $g$ and $B$ leading to an analytical function for the magnetization $M_{dimer}(J,g,B)$ using equation (\ref{mag1}).
With this function $M_{dimer}(J,g,B)$ we performed a combined fit for all temperatures leading to a coupling constant $J$~=~3.5~K and $g$~=~2.04 in excellent agreement with the temperature dependent measurements. The fit results are shown in fig.\ref{MnmvsB} as solid lines. The grey solid line shows the calculated magnetization for a very low temperature $T$~=~0.1~K. Only at this temperature the magnetization steps at the level crossing fields are present. These steps are smoothed even at the lowest measured temperature of 2.4~K due to the thermal averaging process and the small coupling constant J.

The deduced parameter value for $J$ and the $g$-factor are typical for Mn(II) in 
a pseudo-octahedral environment and  oxalate-bridged dimers \cite{Mn-sia-1,Mn-sia-2,Mn-sia-3,Mn-sia-4,Mn-sia-5}.

\section{$\mathrm{Na}_2\mathrm{Co}_2$(C$_2$O$_4$)$_3$(H$_2$O)$_2$}
\subsection{Magnetic Susceptibility}

The Co(II) compound has been synthesized by Price et al. \cite{DJPDalton00}. In SCOX, the Co(II) ions are in the $3d^7$ configuration with an orbitally degenerate ground state $^4T_{1g}$ and a spin of S~=~3/2 per ion in the high-spin state. The presence of an additional orbital moment is already evident from the high temperature Curie-Weiss analysis presented in the introduction. Therefore, a description of the magnetic properties using the spin hamiltonian approach will result in an anisotropic g-factor very different from g=2 and a strong uniaxial anisotropy due to spin-orbit coupling. Alternatively interacting Co(II) ions in octahedral environments are described
by a pseudo spin 1/2 and a strong Ising type anisotropy in the magnetic interaction. Since  the purpose of this work is a comparison of the different spin multiplicities on the transition metal site we use  the model hamiltonian equation (\ref{Ham}).     

The first magnetic susceptibility data on a single crystal parallel and perpendicular to the $a$-axis have been performed by Honda et al. \cite{HondaPRL05}. There the data have been analyzed using a phenomenological S=1/2 two-leg spin ladder function in a limited temperature range only (7 to 20 K). From our systematic study of STOX it is evident that also the Co(II) system can be described as nearly uncoupled dimers of S=3/2 spins on the rungs of the ladder. Therefore we present a new analysis of the single crystal magnetic susceptibility data based on this model similar to the Ni(II) system using hamiltonian (\ref{Ham}). 

\begin{figure}[ht]
    \includegraphics[width=0.9\columnwidth]{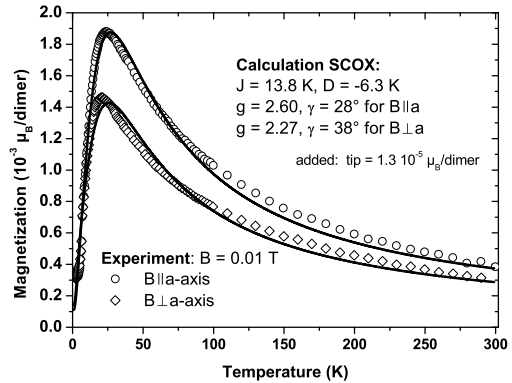}
    \caption{Temperature dependence of the magnetization of a SCOX single crystal at an external field of B$_{ex}$~=~0.01~T along the $a$-axis (circles) and perpendicular to the $a$-axis (diamonds). The data are taken from ref. \cite{HondaPRL05}, the solid lines represent an isolated dimer fit as described in the text.}
		\label{ComvsT}
\end{figure}

A combined fit along both directions (solid lines in fig. \ref{ComvsT}) results in an intradimer coupling constant of $J$~=~13.8~K, an anisotropy of $D$~=~-6.3~K and $g$-values of $g_a$~=~2.60, and $g_{\perp}$~=~2.27~. For comparison with these single crystal measurements, we performed a susceptibility measurement at $B$~=~1~T on the powder sample used for the high field magnetization measurements. An analysis using equation (\ref{totalmag}) with a powder average ($M$~=~($M_\parallel$+2~$M_\perp$)/3) results in $J$~=~11.8~K and $D$~=~-9~K and a large $g$-value anisotropy with $g$-values of $g_{\parallel}$~=~3.90, and $g_{\perp}$~=~1.50~. Note that $J$ is smaller by a factor of $\approx$ 4 than in the Ni(II) system. The variation of the $g$-value is opposite to SNOX due to a different orbital contribution to the total magnetic moment. Since no high field high frequency ESR measurements have been performed so far, the single ion anisotropy strength D had to be used as a free parameter. A satisfactory description of the susceptibility and high field magnetization data is obtained only for a negative value of $D$. In view of the limited accuracy of the single crystal fit the angles to the anisotropy axis  $\gamma_a$~=~28$^\circ$ and $\gamma_{\perp}$~=~38$^\circ$ are consistent with an orientation of the anisotropy axis as in SNOX parallel to the connecting line of the opposite nearest neighbor oxygen ions along the $a$-axis.

\begin{figure}[ht]
    \includegraphics[width=0.9\columnwidth]{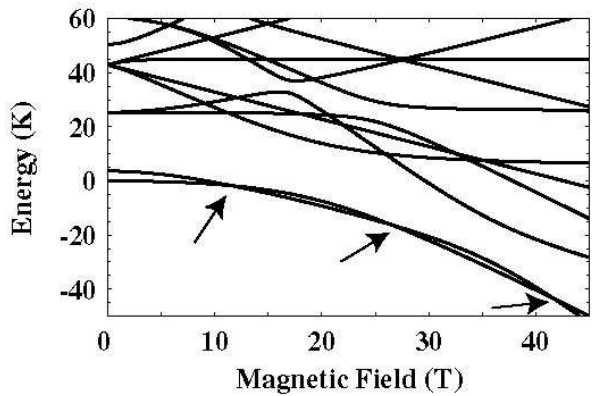}
 \caption{Low energy spin states in SCOX for a magnetic field applied perpendicular to the $z$ axis of the uniaxial
    anisotropy tensor: the $|$m=0$>$ state of the $S\,=\,1$ triplet  and the $S\,=\,0$ singlet are nearly degenerate due to the particular combination of $J$ and $D$. The arrows indicate the
    ground state level crossing fields.}
\label{CoStates}
\end{figure}

Fig.~\ref{CoStates} shows the low energy spin states of SCOX calculated for magnetic fields oriented perpendicular to the main axis of the anisotropy tensor. This orientation gives the dominant contribution in a powder magnetization measurement. For this plot we used the set of  parameters $J$= 11.8~K, $D$=-9~K and $g_\perp$=1.5 derived from the fit of the powder measurements which is also used for the simulation of the high field magnetization measurement as described below. Note the near degeneracy of the $|$m=0$>$ state of the $S\,=\,1$ triplet and the $S\,=\,0$ singlet due to the particular combination of $J$ and $D$. Therefore a series of ground state level crossings is predicted at critical field strengths $B_{Ci}$ of $\approx$ 11.6~T, 26.6~T and 41.7~T in this field orientation.

\subsection{High field magnetization}

\begin{figure}[ht]
    \includegraphics[width=0.9\columnwidth]{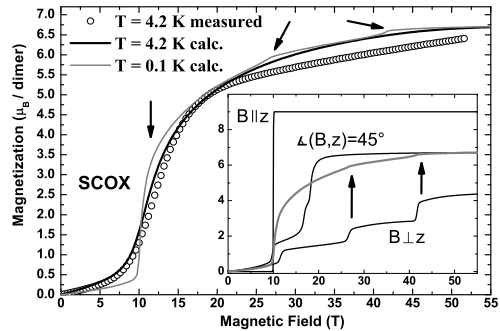}
   \caption{Pulsed high field magnetization of a powder sample of SCOX at 4.2~K.
    The solid lines represent the simulations at 4.2 and 0.1~K. The inset shows single crystal simulations for three different angles between the magnetic field and the anisotropy axis in comparison with the powder average for T=0.1~K.}
		\label{ComvsB}
\end{figure}

We performed a pulsed high magnetic field magnetization measurement at  4.2~K  on a powder sample  of SCOX in magnetic fields up to 52~Tesla (fig.\ref{ComvsB}). It shows one broadened magnetization step from $\approx$ 0.5~$\mu_B$ to nearly 5.5~$\mu_B$ between 10~T and 20~T in addition to a strong finite magnetic susceptibility (gradient in $M(B)$) at low fields. This behavior is in stark contrast to a three step function which can be expected for a dimer of S=3/2 spins from the Breit-Rabi diagram (fig. \ref{CoStates}). In this case when the antiferromagnetic exchange is of similar strength as the single-ion anisotropy it is necessary for a  qualitative and quantitative description of a powder magnetization measurement to perform a  weighted average over a full set of angles between the anisotropy axis and the field direction from 0$^\circ$ to 90$^\circ$ in steps of 1$^\circ$ with varying g from $g_\parallel$ to $g_\perp$ since the Breit-Rabi diagram depends strongly on this angle. This is illustrated in the inset of fig. \ref{ComvsB} where we plot the calculated single crystal magnetization curves at $T$ = 0.1~K for three different angles in comparison with the powder average. Due to the strong anisotropy for B $||$ z the first level crossing at 10~T leads from S~=~0 to the
% Neu:
 $|$m=-3$>$ of the S~=~3 fully polarized sextet
 state resulting in a sharp step of the magnetization curve. This situation contributes significantly to the powder average.  
The black solid line in fig.~\ref{ComvsB} is calculated for $T$= 4.2~K  using the set of parameters obtained by the powder susceptibility with an optimized  value $g_\parallel$=3.0. Due to the dominance of the large step of the $B || z$ direction, the influence of the steps in the $B \perp z$ direction in the powder sample is noticeable only at very low temperatures. Note that in fig.\ref{ComvsB}, at T~=~0.1~K the magnetization shows two small kinks at the level crossing fields of the $B \perp z$ direction (arrows on grey solid line). These kinks result from the powder averaging which leads to a smoothing of the steps (cf. inset of fig.\ref{ComvsB}). Recently we became aware of experiments on Co-Oxalate by Y. Nakagiwa et al. which qualitatively agree with our findings \cite{nakagawa06}.

In the literature a very wide range of values for $D$ and the $g$-factor are found for Co(II) complexes. See e.g. the review by R. Boca \cite{Boca04} or examples in \cite{Co-sia-1,Co-sia-2,Co-sia-3,Co-dim-1,Co-dim-2,Co-dim-3,Co-dim-4}.

\section{$\mathrm{Na}_2\mathrm{Fe}_2$(C$_2$O$_4$)$_3$(H$_2$O)$_2$}

\subsection{Magnetic Susceptibility}

The magnetic susceptibility of a powder sample of the Fe(II) compound  has been measured by Kreitlow et al. \cite{KreitlowJMMM04}. In SIOX, the Fe(II) ions are in the $3d^6$ configuration with an orbitally degenerate ground state  $^5T_{2g}$ and a spin of S~=~2 per ion in the high-spin state. The presence of an additional orbital moment is already evident from the high temperature Curie-Weiss analysis of the magnetic susceptibility. Comprehensive single crystal  magnetic susceptibility data parallel and perpendicular to the $a$-axis have been published by Kikkawa et al. \cite{KikkawaJPSJ05} showing a very strong anisotropy. Only the data parallel to the $a$-axis have been analyzed using a singlet-triplet system including a uniaxial anisotropy resulting in a singlet-triplet gap of 51~K and a very large anisotropy D=41~K. 
Here we present a new analysis of the single crystal magnetic susceptibility data parallel and perpendicular to the $a$-axis based on a dimer model using hamiltonian (\ref{Ham}). 

\begin{figure}[ht]
    \includegraphics[width=0.9\columnwidth]{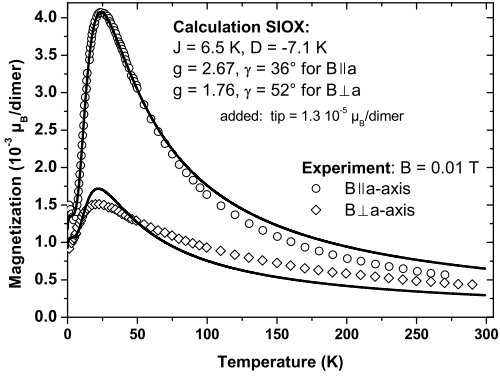}
    \caption{Temperature dependence of the magnetization of a SIOX single crystal at an external field of B$_{ex}$~=~0.01~T along the $a$-axis (circles) and perpendicular to the $a$-axis (diamonds). The data are from ref. \cite{KikkawaJPSJ05}, the solid line represents an isolated dimer fit as described in the text.}
		\label{FemvsT}
\end{figure}

The combined fit along both directions (solid lines in fig. \ref{FemvsT}) results in an intradimer coupling constant of $J$~=~6.5~K, $g$-values of $g_a$~=~2.67, and $g_{\perp}$~=~1.76~. Note that $J$ is reduced by a factor of $\approx$ 2 with respect to the Co(II) system. The single ion anisotropy strength D has been determined to $D$=-7.1~K. 
Similar to the SCOX compound, we performed magnetic susceptibility measurements on a SIOUX powder sample used for the high field magnetization measurements and analysed these data in the same way. This results in a parameter set of $J$~=~6.7~K, $D$~=~-~8.7~K, $g_\parallel$~=~2.67, and $g_{\perp}$~=~2.65~ which we used to calculate the Breit-Rabi diagram (see inset of fig. \ref{FemvsB}). Therefore in this system the isotropic magnetic exchange and the single ion anisotropy are of similar size leading to the strong anisotropy in the measured magnetic susceptibility. In view of the limited accuracy of the fit the angles to the anisotropy axis  $\gamma_a$~=~35$^\circ$ and $\gamma_{\perp}$~=~52$^\circ$ are again consistent with an orientation of the anisotropy axis  parallel to the connecting line of the opposite nearest neighbor oxygen ions along the $a$-axis.

\subsection{High field magnetization}

\begin{figure}[ht]
    \includegraphics[width=0.9\columnwidth]{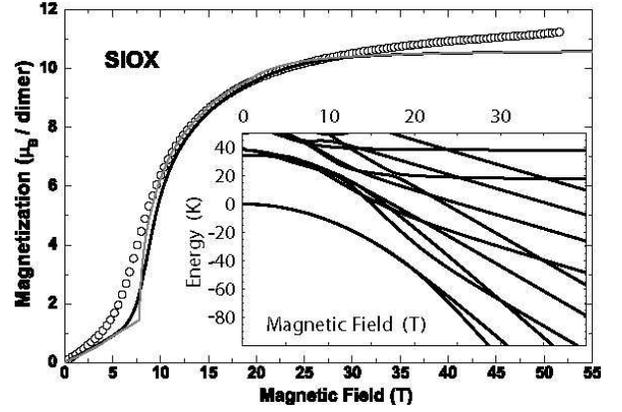}
   \caption{Pulsed high field magnetization of a powder sample of SIOX at 4.2~K.
    The solid lines represent simulations at 4.2 and 0.1~K. The inset shows the low energy states 
    in a Breit-Rabi diagram for a magnetic field applied perpendicular to the $z$ axis of the uniaxial
    anisotropy tensor.}
		\label{FemvsB}
\end{figure}

We performed a pulsed high magnetic field magnetization measurement at 4.2~K on a powder sample of SIOX in magnetic fields up to 52~Tesla (fig.\ref{FemvsB}). It shows the same oveall features as the experiment on the Co analogue, namely a broadened magnetization step from $\approx$ 1.0~$\mu_B$ to nearly 10~$\mu_B$ per dimer between 7 and 20~T in addition to a strong finite magnetic susceptibility at low fields.  The low energy part of the Breit-Rabi diagram for B perpendicular to the anisotropy axis using a set of parameters $J$= 6.7~K, $D$=-8.7~K, 
and $g_\perp$=2.65 optimized to describe the high field magnetization data is shown in the inset of fig.\ref{FemvsB}. The $|$m=0$>$ state of the $S\,=\,1$ triplet  and the $S\,=\,0$ singlet are very close and several level crossings occur between 5 and 40~T.

In the literature similar to Co(II)  a very wide range of values for $D$ and the $g$-factor is found for Fe(II) complexes. See e.g. the review by R. Boca \cite{Boca04} or examples in \cite{Fe-sia-1,Fe-sia-2,Fe-sia-3,Fe-dim-1,Fe-dim-2,Fe-dim-3,Fe-dim-4,Fe-dim-5}.

\section{Magnetic exchange pathways}
\label{electronic-structure}

An unexpected result of the present study is the very large difference between the strength 
of the magnetic exchange interaction along the rungs  and legs of the spin ladder
leading to the applicability of the dimer model in this system.
This can be qualitatively understood considering the different coordination and superexchange angles
of the exchange mediating oxalate molecules. 

\begin{figure}
    \includegraphics[width=0.9\columnwidth,clip]{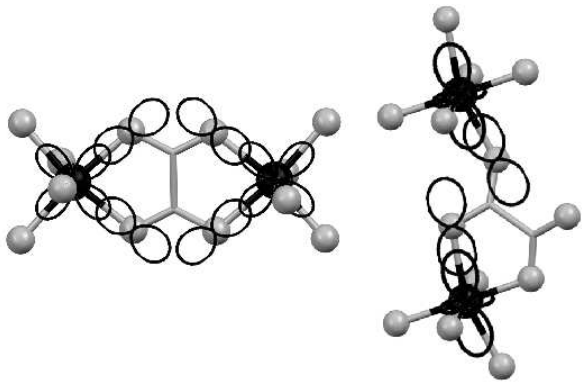} 
    \caption{ T-(ox)-T coordination and proposed magnetic exchange pathways on the rungs (left panel) and  on the legs of the spin ladder (right panel). The involved T 3d and oxygen 2p orbitals are shown in black. }
\label{coordination}
\end{figure}

On the rungs the oxalate forms a $\mu$-1,2,3,4 bridge between two transition metal ions. This bridge provides two symmetric superexchange pathways. On each path the transition metal 3d x$^2$-y$^2$ orbitals have an enhanced electron probability density extending directly towards the corresponding oxygen ions. There is a direct overlap of the transition metal 3d x$^2$-y$^2$ and the O 2p wave functions forming $\sigma$ bonds. The polarized 2p orbitals themselves are strongly overlapping. Therefore the intermediate carbon atom is not involved in the superexchange mechanism   resulting in a strong antiferromagnetic superexchange interaction (see fig. \ref{coordination}). In the literature several $\mu$-1,2,3,4 oxalato-bridged Ni(II) dimer and chain systems are reported  (see e.g. \cite{escuerICA95,romanIChem96} and references therein) with $J$ values in the range of 20 to 42~K.

For the  oxalate bridge along the legs of the spin ladder the situation is completely different.
In this case a $\mu$-1,2,3 oxalato bridge is formed. As shown in the right panel of fig. \ref{coordination} only one transition metal ion forms two covalent bonds (one with the x$^2$-y$^2$ orbital and one with the 3z$^2$ - r$^2$ orbital) with  the oxalate molecule whereas of the second transition metal ion only the 3z$^2$ - r$^2$ orbital forms one covalent bond. The O 2p orbitals involved in the T-O bonds are not overlapping with each other and therefore either the carbon atoms or  orthogonal O 2p orbitals are involved  which strongly suppresses the strength of the superexchange mechanism and may even lead to a weak ferromagnetic
exchange. Similar $\mu$-1,2,3 oxalato bridges are reported for Cu(II) systems with magnetic exchange strengths in the range of -0.2 to 0.3~K \cite{nunezICA01,castilloICC01} consistent with the estimate of $|K|$ \cite{MennerichPhyRevB06} being negligibly small in STOX.

\begin{figure}
    \includegraphics[width=0.9\columnwidth,clip]{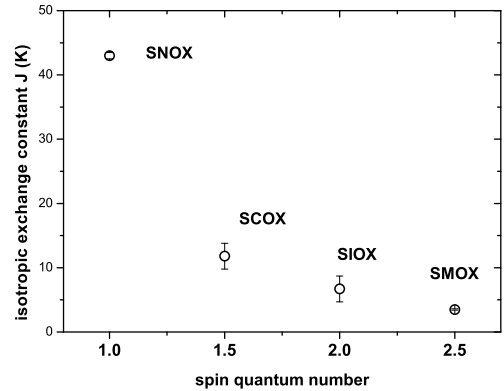} 
    \caption{ Change of the isotropic magnetic exchange constant $J$ with decreasing number of 3d electrons, i.e. increasing spin quantum number, in the series Na$_2$T$_2$(C$_2$O$_4$)$_3$(H$_2$O)$_2$.}
\label{Jseries}
\end{figure}

A second interesting result is the continuous decrease of the isotropic magnetic exchange strength $J$ going from 3d$^8$ (Ni(II)) to 3d$^5$ (Mn(II)) shown in fig. \ref{Jseries}. 
The same systematics is present e.g. in the series of three dimensional antiferromagnets
KTF$_3$ with T=Ni,Co, Fe, Mn \cite{Dovesi-FD97}. There the nearest neighbor exchange varies
from 44-50~K in KNiF$_3$ over 19~K in KCoF$_3$ and 6~K in KFeF$_3$  to 3.6~K in KMnF$_3$.
This systematics can be regarded as a result of the description of the total magnetic exchange energy in terms of the product $J\, \mvec{S}_1 \mvec{S}_2$. As described above the 3d x$^2$-y$^2$ orbital contributes the most to the total spin exchange energy. In Co(II), Fe(II) and Mn(II) the additional spin density is located in the t$_{2g}$ orbitals. These orbitals do not point directly towards the oxygen ions. They may contribute only very little to the
magnetic exchange energy but increase the total spin value. As a consequence the numerical value of the exchange constant $J$ is reduced. 

\section{Conclusion}

In this paper we demonstrate the application of high magnetic fields to study the magnetic properties of low dimensional spin systems. We present a case study on the series of 2-leg spin-ladder compounds Na$_2$T$_2$(C$_2$O$_4$)$_3$(H$_2$O)$_2$ with T~=~Ni, Co, Fe and Mn. In all compounds the transition metal is in the 2+ high spin configuation. The localized spin varies from S=1 to 3/2, 2 and 5/2 within this series. 
The magnetic properties were examined experimentally by magnetic susceptibility, high field ESR, pulsed high field magnetization  and specific heat measurements. The data are analysed using a spin hamiltonian description with Heisenberg exchange interaction. Although the transition metal ions form structurally a 2-leg ladder, an isolated dimer model consistently describes the observations very well. All compounds exhibit magnetic field driven ground state changes which at very low temperatures lead to a multistep behaviour in the magnetization curves. In the Co and Fe compounds a strong axial anisotropy induced by the orbital magnetism leads to a nearly degenerate ground state and a strongly reduced critical field. We find a monotonous decrease of the intradimer magnetic exchange if the number of 3d electrons is decreased which indicates
that the additional spin density in the t$_{2g}$ orbitals does not contribute to the exchange energy.

\section{Acknowledgments}
Financial support by the Deutsche Forschungsgemeinschaft through SPP~1137 ''Molecular Magnetism'' grant KL 1086/6-1 and 6-2 is gratefully acknowledged. The work of R.K. in
Toulouse was supported by the DFG through grant KL 1824/1-1. 
D.J.P. is grateful to the EPSRC of the UK for the award of an Advance Research Fellowship (Gr/A00836/02).

\end{document}